\documentclass{MBE}
\usepackage{amsmath}
 \usepackage{paralist}
  \usepackage{graphics}
   \usepackage{epsfig}
    \usepackage[colorlinks=true]{hyperref}

 \hypersetup{urlcolor=blue, citecolor=red}

\def\currentvolume{xx}
 \def\currentissue{xx}
  \def\currentyear{20xx}
   \def\currentmonth{xx}
    \def\ppages{1--xx}

\theoremstyle{definition}

\newcommand{\ud}{\mathrm{d}}

\newcommand{\HH}{\mathcal{H}}
\begin{document}
\title[Spatial Model of Chemotherapy]{A Spatial Model of Tumor-Host Interaction: Application of Chemotherapy}

\thanks{$^\dagger$These authors contributed equally to this paper.}
\thanks{$^\ast$To whom correspondence should be addressed. Phone: +1 612 626-1307, e-mail: {\tt hinow@ima.umn.edu}}
\maketitle
\centerline{\scshape Peter Hinow$^{\dagger\,\ast}$}
{\footnotesize
\centerline{Institute for Mathematics and its Applications}
\centerline{University of Minnesota, 114 Lind Hall, Minneapolis, MN 55455, USA}
} 
\medskip
\centerline{\scshape Philip Gerlee$^\dagger$}
{\footnotesize
\centerline{Niels Bohr Institute, Center for Models of Life, Blegdamsvej 17, 2100 Copenhagen, Denmark}
} 
\medskip
\centerline{\scshape  Lisa J.~McCawley, Vito Quaranta}
{\footnotesize
\centerline{Department of Cancer Biology, Vanderbilt University, Nashville, TN 37232, USA}
} 
\medskip
\centerline{\scshape  Madalina Ciobanu}
{\footnotesize
\centerline{Department of Chemistry, Vanderbilt University, Nashville, TN 37235, USA}
} 
\medskip
\centerline{\scshape  Shizhen Wang}
{\footnotesize
\centerline{Department of Surgical Research}
\centerline{Beckman Research Institute of City of Hope, Duarte, CA 91010, USA}
} 
\medskip
\centerline{\scshape Jason M. Graham, Bruce P.~Ayati}
{\footnotesize
\centerline{Department of Mathematics, University of Iowa, Iowa City, IA 52242, USA}
} 
\medskip
\centerline{\scshape Jonathan Claridge, Kristin R. Swanson}
{\footnotesize
\centerline{Department of Applied Mathematics and Department of Pathology}
\centerline{University of Washington, Seattle, WA 98195, USA}
} 
\medskip
\centerline{\scshape Mary Loveless}
{\footnotesize
\centerline{Department of Biomedical Engineering, Vanderbilt University, Nashville, TN 37232, USA}
} 
\medskip
\centerline{\scshape Alexander R.~A.~Anderson}
{\footnotesize
\centerline{H.~Lee Moffitt Cancer Center \& Research Institute, Integrated Mathematical Oncology,}
\centerline{12902 Magnolia Drive, Tampa, FL 33612, USA}
} 
\medskip
\centerline{(Communicated by James Glazier)}
\begin{abstract}
In this paper we consider chemotherapy in a spatial model of tumor growth. The model, which is of reaction-diffusion type, takes into account the complex interactions between the tumor and surrounding stromal cells by including densities of endothelial cells  and the extra-cellular matrix. When no treatment is applied the model reproduces the typical dynamics of early tumor growth. The initially avascular tumor reaches a diffusion limited size of the order of millimeters and initiates angiogenesis through the release of vascular endothelial growth factor (VEGF) secreted by hypoxic cells in the core of the tumor. This stimulates endothelial cells to migrate towards the tumor and establishes a nutrient supply sufficient for sustained invasion. To this model we apply cytostatic treatment in the form of a VEGF-inhibitor, which reduces the proliferation and chemotaxis of endothelial cells. This treatment has the capability to reduce tumor mass, but more importantly, we were able to determine that inhibition of endothelial cell proliferation is the more important of the two cellular functions targeted by the drug. Further, we considered the application of a cytotoxic drug that targets proliferating tumor cells. The drug was treated as a diffusible substance entering the tissue from the blood vessels. Our results show that depending on the characteristics of the drug it can either reduce the tumor mass significantly or in fact accelerate the growth rate of the tumor. This result seems to be due to complicated interplay between the stromal and tumor cell types and highlights the importance of considering chemotherapy in a spatial context.
\end{abstract}

\section{Introduction}\label{Intro}

Cancer is a complex multiscale disease and although advances have been made in several areas of research in recent years the dynamics of tumor growth and invasion are still poorly understood. It has been recognized that tackling this problem requires the collaborative effort of scientists from several disciplines such as genetics, cell biology, biological physics and mathematical biology to name but a few. This paper introduces a mathematical model of tumor invasion that was developed in precisely such a multi-disciplinary effort. Far from claiming that this type of collaboration is groundbreaking, we would still like to highlight the potential behind such a inter-disciplinary collaboration in cancer research. 

The second annual workshop of the Vanderbilt Integrative Cancer Biology Center (VICBC) took place from July 18-21, 2006 at Vanderbilt University in Nashville, Tennessee. During that workshop, our group of $16$ biologists, mathematicians and biomedical engineers developed a mathematical model for tumor invasion. The model focuses on the complex interactions between different types of cells (normoxic, hypoxic, endothelial cells etc.) and their interaction with the surrounding extracellular matrix. 
The mathematical model is fully continuous with respect to time and space and consists of partial differential equations of reaction-diffusion type. The aim of this paper  is to study cancer chemotherapy in a spatial context. 

The plan of the paper is as follows: First we discuss the biological background of the problem and discuss several aspects of tumor invasion and therapy including previous mathematical modelling. We then introduce our model and present numerical simulation results. We discuss our results in the context of other mathematical models and clinical applications and we outline possible directions for future research. An appendix describes the non-dimensionalization of the model.

\section{Biological background}\label{Bio}

\subsection{Cell proliferation and survival}
Cancer is regarded as a group of diseases characterized by abnormal cell growth and division, leading to tumors and eventually metastasis.  When the disease is not present, the healthy cells in the tissue have the ability to proliferate and nutrients are kept at normal levels. They follow an ordinary cell cycle and upon growth and division create an increased cell density; when a certain threshold is met, the cells push on each other, creating a ``crowding effect.'' In this case, normal cells eventually become quiescent or may even go into apoptotis (programmed cell death). However, when normal cells become tumorigenic, the cells will grow into three dimensional  structures to overcome some of the crowding effect.  The healthy balance between normal and apoptotic cells is disrupted and inside the tumor arises a new type of cell capable of sustaining itself with a lower nutrient level than the normal cells: the hypoxic cell. One of the most important roles played by hypoxic cells within a tumor is their ability to recruit new vasculature \cite{Brahimi} to provide an additional supply of nutrients, such as oxygen, glucose, and growth factors. Oxygen is one of the most important nutrients in the study of cancer. It has been widely studied and documented with respect to its levels for various tumors due to the ease of measurement, and we have thus used it as our nutrient of choice for the work presented here.

\subsection{Unbiased and biased cell migration}

One of the key aspects of cancer invasion is cell motility. Tumor cells can move within their environment by both undirected and directed mechanisms.  Unbiased cell migration maintains no recognizable pattern to the movement while cells undergoing biased migration follow a particular pattern, generally along a gradient, driven by physical or chemical means. This mathematical model will describe both types of movement using four mechanisms of motility. Diffusion, haptotaxis, crowding (as driven by tumor and non-tumor cells), and chemotaxis (as observed in angiogenesis).

Diffusion is unbiased cell movement similar to the phenomenon of Brownian motion.  This type of movement describes each cell's ability to randomly move within the surrounding environment.  This random motility is applicable to all cell types included within our model. As the tumor becomes progressively more invasive, the tumor cells migrate toward dense areas of extracellular matrix (ECM) \cite{Aznavoorian, McCarthy}. This directed movement towards higher density of ECM proteins is known as haptotaxis. 

As the tumor cells proliferate, the tumor expands outward. As pressure increases from this proliferation, the cells are pushed into a new area \cite{DeAngelis2000,Gurtin1977}.  This pressure-driven movement is described as crowding. An example of this crowding effect involves colorectal cancers. Adenomatous tissue forms and invades within the villi of intestinal epithelium. As the space within the villi fill, the villus epithelium tears, and the adenomous tissue extends into the gut \cite{Bienz}. The tumor cells will continue to fill available space until crowding creates enough pressure to cause compression or damage to surrounding tissue; thus, crowding perpetuates the aggression of the tumor.

\subsection{Hypoxia-driven angiogenesis}

As the cells proliferate and the tumor grows in size, nutrients and oxygen become  the most critical factors for cell viability. When a small avascular tumor exceeds a certain size (with a critical diameter of approximately $2\, mm$), simple diffusion of oxygen to highly metabolizing tissues becomes inadequate, and hypoxia will be induced in the cell population lacking nutrients. This will, in turn, trigger the production and release of numerous tumor angiogenic factors (TAFs) to stimulate angiogenesis \cite{Folkman}. Pathological angiogenesis is hence a hallmark of cancer. Vascular endothelial growth factor (VEGF) is the most  prominently induced molecule and plays a predominant role in angiogenesis \cite{Ferrara03}. This process not only affects the blood supply to meet the metabolic demands of the growing mass of cells, but also represents the track to deliver anti-cancer drugs to all regions of a tumor in effective quantities \cite{Carmeliet}.

\subsection{Chemotherapy}

Chemotherapy technically refers to the use of drugs that selectively target the cause of a disease, but is now almost synonymous with the use of drugs to treat cancer. The mechanisms of these drugs are very diverse, but they usually target cell proliferation by for example damaging the DNA or disrupting the mitotic spindles, which are essential for successful cell division. Chemotherapeutic drugs can be divided into three classes. Class I drugs are non-phase specific and affect both quiescent and proliferating cells; class II drugs are phase-specific and only target cells in a given phase; and finally class III drugs only kill cells that are proliferating. One example of a class III drug is Docetaxel  \cite{lyseng05}, which induces cell death during mitosis by stabilizing micro-tubules and thus hindering the formation of functional mitotic spindles. In this paper we will focus on the modelling of class III drugs and investigate how the characteristics of the drug influences the efficacy of the treatment.

The drugs mentioned so far are considered cytotoxic as they actively kill tumor cells. Another family of drugs we will consider are cytostatic drugs, which act by inhibiting cell division or some specific cell function in the cancer cells or host cells directly involved in tumor invasion. One example of this type of drug is Tamoxifen \cite{jordan06} used in the treatment of breast cancer, which binds to estrogen receptors on the cancer cells and therefore inhibits transcription of estrogen-responsive genes. Another example is Avastin \cite{adams05}, which does not target the cancer cells directly but instead affects the surrounding endothelial cells reducing the formation of new blood vessels supplying the tumor with nutrients. Finally, some drugs have been shown to have both cytostatic and cytotoxic effects, such as Lapatinib \cite{hinow07}.

\section{Mathematical models of tumor growth and treatment}\label{Overview}

\begin{sloppypar}
Many researchers have proposed mathematical models for tumor growth and treatment. Let us mention here \cite{Anderson05,DeAngelis2000,Anderson06,Chaplain06, Ribba, Bertuzzi, gerlee07} among others. A recent book in the Springer Lecture Notes in Mathematics series \cite{Friedman} contains further interesting survey articles. In these works the authors have investigated phenomena such as invasion, angiogenesis, and cytotoxic chemotherapy. 
\end{sloppypar}

Mathematical models have considered random motility as a major mechanism of cellular motion \cite{Murray}. With continuous diffusion well-understood mathematically, this assumption forms the backbone of numerous models, including the majority of those that will be mentioned here.  Random motility has been studied in the context of gliomas by Swanson \cite{Swanson03, Harpold07}, who has done extensive comparisons with patient data. It has also been used to motivate the probabilistic movement of individual cells in models by Anderson and Chaplain \cite{Anderson98} and Anderson \cite{Anderson05}. Random motility is especially convenient because it is mathematically identical to the diffusion that governs chemical motion.  This allows the inclusion of chemical species, such as nutrients and tumor angiogenic factors, while remaining in the coherent mathematical framework of reaction-diffusion equations \cite{Hundsdorfer}.

The model presented in this paper examines cell types in terms of spatially-dependent cell densities, and chemical species in terms of spatially-dependent concentrations. This does not permit analysis of small-scale structures, particularly the vascular networks simulated in many angiogenesis models \cite{Anderson98, Anderson05, Chaplain00, Chaplain06, Jones06}.  However, it does allow for a large-scale perspective, and gives us a broader range of options in the interactions we consider.

Tumor-induced angiogenesis has been the subject of considerable attention in mathematical models.  Tumor angiogenic factor (TAF) plays a central role in driving such models, providing the means for a growing tumor to initiate the growth of new vasculature.  The initial model of Chaplain and Stewart \cite{Chaplain91} uses a free-boundary problem to describe the interaction of TAF and endothelial cells implicitly.  Since then, however, chemotaxis has become the normal way of modelling this interaction, with endothelial cells moving up the gradient of TAF concentration \cite{Chaplain93, Anderson98, Anderson05, Orme97}.  A number of models also include haptotaxis of endothelial cells up a gradient of extracellular matrix \cite{Anderson98, Chaplain00, Chaplain06, Orme97}. We have focused on chemotaxis as the dominant mechanisms of endothelial cell movement.

Hypoxia is widely believed to be crucial to angiogenesis, with hypoxic cells secreting TAF. A number of models have considered the transition of tumor cells between various states, such as Adam and Megalakis \cite{Adam90} and Michelson and Leith \cite{Michelson97}, and the diffusive model of Sherratt and Chaplain \cite{Sherratt01}. But in these models hypoxia is only included in the sense of quiescent, non-proliferating cells. On the other hand, vessel recruitment stimulated by TAF-secreting hypoxic cells has been modelled in a hybrid-discrete setting \cite{alarcon06}.

Chemotherapy has been subject to extensive mathematical modelling. These efforts have mainly focused on drug resistance \cite{abundo89, costa95, usher96}, cell cycle specific drugs using age-structured populations \cite{gaffney04, hinow07}, and optimal treatment scheduling \cite{barbolosi2001, murray90, panetta2003}. Most previous models of chemotherapy have neglected the spatial effects and therefore disregarded an important component in tumor growth. Low levels of nutrient within the tumor can cause a large fraction of the tumor cell population to be quiescent and these cells are less responsive to chemotherapy. Because this is a spatial effect caused by the diffusion of nutrients from the blood vessels to the tumor, we believe that it is important to include space in a mathematical model of chemotherapy. This has been investigated in both a hybrid-discrete approach \cite{alarcon06} and a continuous setting \cite{norris06}, but the model presented in this paper is different in that it takes into account the tumor-host interactions in a more detailed manner.	

\section{The mathematical model}\label{MathMod}
Our mathematical model is based on partial differential equations for the densities of all cells and chemical factors involved. Thus it is fully continuous with respect to time and space. Let $t$ and $x$ denote time and space, respectively.  The complete set of dependent variables are listed in Table \ref{dependentVar}.

The density of all types of cells and matrix combined is
\begin{equation*}
v= h+a+n+f+m.
\end{equation*}

The equation for oxygen concentration is
\begin{equation}\label{nutrient}
\frac{\partial w}{\partial t}(x,t) =  D_w \frac{\partial^2 w}{\partial x^2} +\alpha_w m (w_{max}-w) - \beta_w(n+h+m)w - \gamma_w w,
\end{equation}
where $D_w$ is the coefficient of oxygen diffusion, $w_{max}$ is the maximum oxygen density, and $\beta_w$ is the uptake rate of oxygen by normoxic, hypoxic and endothelial cells. Oxygen is a diffusible substance that is provided at a rate $\alpha_w$ by the vasculature, whose amount in the unit volume is assumed to be proportional to the endothelial cell density. This modelling approach is found for instance in \cite{Anderson05}. The reason for the choice of the source term $\alpha_w w (w_{max}-w)$ is that, at high environmental levels of oxygen, less oxygen is released through the vessel walls (i.e. by the vasculature). Finally oxygen decays at a rate $\gamma_w$.

The normoxic cells are governed by
\begin{align}
\frac{\partial n}{\partial t}(x,t)&=  \frac{\partial }{\partial x}\left( (D_n\max\{n-v_c,0\}+D_m)\frac{\partial n}{\partial x}\right) - \frac{\partial }{\partial x}\left(\chi_n
 n\frac{\partial f}{\partial x} \right) \label{normoxic} \\
&\quad+\alpha_nn(v_{max}-v) -\alpha_h\HH(w_h-w)n + \frac{1}{10}\alpha_h\HH(w-w_h)h. \nonumber
\end{align}
Normoxic cells possess a ``background'' random motility $D_m$. For simplicity, we assume that this  random motility is constant, although recent work by Pennacchietti et al.~\cite{Pennacchietti2003} indicates that cells under hypoxic conditions can become more mobile. On top of this, a concentration of normoxic cells above a threshold $v_c$ adds to the dispersion of these cells through crowding-driven motion represented by the nonlinear diffusion term $D_n \max\{n-v_c,0\} n_x$. Such a pressure driven motility was first proposed in a paper by Gurtin and MacCamy \cite{Gurtin1977}.  On the first line of equation \eqref{normoxic} we find the term responsible for the haptotactic movement up a gradient of extracellular matrix. The haptotactic coefficient is denoted by $\chi_n$. The second line describes the gain and loss terms for normoxic cells. First, we assume a logistic growth with rate $\alpha_n$. The growth levels off in regions where the sum of all cells and matrix approaches the maximal density $v_{max}$. In those regions where the concentration of oxygen drops below a certain critical value $w_h$, normoxic cells enter the hypoxic class at a rate $\alpha_h$. Here $\HH$ denotes the Heaviside function which is $1$ for positive arguments and zero otherwise. The transition process is reversible, and hence there is an influx from the hypoxic class at a reduced rate $\frac{1}{10}\alpha_h$ (to our knowledge, this reduction factor is not available in the literature, we choose the value $\frac{1}{10}$). We assume that it takes some time for a cell to resume the cell cycle after it has been recruited from a quiescent state. Note that the two processes are mutually exclusive.

The equation for hypoxic cells is
\begin{equation}\label{hypoxic}
\frac{\partial h}{\partial t}(x,t)= \alpha_h\mathcal{H}(w_h-w)n - \frac{1}{10}\alpha_h\mathcal{H}(w-w_h)h
-\beta_h\mathcal{H}(w_a-w)h.
\end{equation}
The first and the second terms are dictated by conservation of mass and correspond to terms in equation \eqref{normoxic}. The third term describes the transition of hypoxic cells to apoptotic cells at rate $\beta_h$ as the level of oxygen falls below a second threshold, $w_a<w_h$. Hypoxic cells are less active in general due to reduced availability of oxygen and other nutrients and we assume that lack of energy causes them to be immobile.

The equation for apoptotic cells is given by
\begin{equation}\label{apoptotic}
\frac{\partial a}{\partial t}(x,t)= \beta_h\mathcal{H}(w_a-w)h.
\end{equation}
The first term corresponds to the third term in equation \eqref{hypoxic}. 

The equation for endothelial cells is
\begin{equation}\label{endothelial}
\frac{\partial m}{\partial t}(x,t)= \frac{\partial }{\partial x}\left( D_m\frac{\partial m}{\partial x} -m \chi_m \frac{\partial g}{\partial x}\right) +\alpha_m mg(v_{max}-v). \end{equation}
For simplicity and to reduce the number of free parameters we assume endothelial cells possess the same random motility $D_m$ as normoxic cells. Endothelial cells respond via chemotaxis to gradients of angiogenic factor $g$ and require the presence of angiogenic factor for proliferation. Proliferation is capped by the total density of cells. The proliferation constant for endothelial cells is $\alpha_m$. In a previous version of the model we also incorporated a haptotaxis term (in response to gradients of extracellular matrix), however, this made little difference to the simulation outcomes discussed here. We therefore decided, for the sake of simplicity, to omit haptotaxis from the endothelial cell equation. The proliferation of endothelial cells is governed by the same volume exclusion constraint as the proliferation of normoxic cells.

The equation for the extracellular matrix is 
\begin{equation}\label{tissue}
\frac{\partial f}{\partial t}(x,t)= - \beta_f n f.
\end{equation}
Tissue matrix is degraded by the tumor cells according to a mass-action law with rate constant $\beta_f$. 

The equation for angiogenic factor (VEGF) is
\begin{equation}\label{vegf}
\frac{\partial g}{\partial t}(x,t)= D_g\frac{\partial^2 g }{\partial x^2}+ \alpha_g h - \beta_g m g.
\end{equation}
Angiogenic factor moves by standard linear diffusion with coefficient $D_g$. Angiogenic factor is produced by hypoxic cells alone at rate $\alpha_g$, taken up by endothelial cells with a mass-action coefficient of $\beta_g$. Endothelial growth factor is a survival factor as well, so the endothelial cells consume VEGF at all times, not just when they proliferate.

This system of equations will represent our baseline model, but in order to include therapy with a cytotoxic drug we will extend our model in section \ref{cytotox_therapy}. In the next section we will non-dimensionalize the model and discuss values for the parameters.

\section{Parametrization}\label{Nondim}
In order to simplify the analysis and simulations of the model we non-dimensionalize the model using the characteristic scales shown in Table \ref{tab_scaling}. Although key parameters in our model, such as the oxygen consumption rates and diffusion constants, have been reported in the literature, the non-dimensionalization makes it easier to estimate parameters that have not been experimentally determined by comparing the influence of different processes to each other. A helpful article in this respect is \cite{Segel}. The detailed non-dimensionalization is carried out in the appendix.

The oxygen diffusion constant was set to $D_w=10^{-5}$ cm$^2$ s$^{-1}$, as in \cite{Anderson05}. The oxygen consumption of cancer cells has been experimentally determined to $\beta_w = 6.25\times 10^{-17}$ mol cell$^{-1}$ s$^{-1}$ \cite{Casciari92}, and for simplicity we assume that all cell types have this consumption rate. The oxygen production rate is difficult to estimate as it depends on vessel permeability and the oxygen concentration in the blood. We therefore make a non-dimensional estimate and set it to $\alpha_w=1$, of the same order as the consumption rate. The oxygen decay rate is set to the non-dimensional value $\gamma_w=0.025$ \cite{Anderson05}.

The random cell motility parameter $D_m = 10^{-9}$ cm$^2$ s$^{-1}$ has been estimated \cite{Bray} and for simplicity we assume that the pressure driven motility $D_n$ is of the same magnitude. The density at which this motility occurs is estimated to be 80\% of the maximum density $v_{max}$.
Parametrizing the haptotactic coefficient is difficult as we consider a matrix consisting of a variety of cells and macro-molecules (MM), while the haptotactic movement of cancer cells is only sensitive to certain molecules bound within the matrix. From experimental data we know that the concentration of these molecules within the matrix is in the range $10^{-8}-10^{-11}$ M \cite{Terranova}. The dimensional haptotactic parameter has been estimated to $\chi \sim 2600$ cm$^2$ s$^{-1}$ M$^{-1}$ \cite{Anderson98}, and using this value we can at least give a non-dimensional estimate to the haptotaxis coefficient $\chi_n=1.4 \times 10^{-4}$ \cite{Anderson05}. 

By rescaling time by the average doubling-time of the cancer cells the growth rate is by definition $\alpha_n= \log 2$. The conversion rate from normoxic to hypoxic cells at low oxygen pressure has to our knowledge not been experimentally determined, but a reasonable guess is that this occurs on the time-scale of hours. The rate is therefore estimated at $\alpha_h=2.8 \times 10^{-5}$ s$^{-1}$, which means that in hypoxic conditions half of the cells will turn hypoxic within $\log 2/2.8 \times 10^{-5} \approx$ 7 hours. The process of apoptosis is on the time-scale of minutes \cite{tyas00}, but the time it takes before the cells commits to this decision is much longer and on the order of days \cite{borutaite05}. The conversion rate from hypoxic to apoptotic cells is therefore estimated to be considerably smaller and set to $\beta_h=5.6\times10^{-6}$ s$^{-1}$. The exact oxygen concentrations at which the above conversions occur is difficult to measure as it depends not only on the cell type, but also on other environmental factors such as acidity and the concentration of various growth factors. Instead these parameters can be estimated from oxygen concentrations measured in the necrotic centers of real tumors. Such measurements show that the oxygen concentration is approximately 0.5-30 \% of that in the surrounding tissue \cite{brown04}. From this we estimate the hypoxic threshold $w_h$ to be 5 \% of the background oxygen concentration and the apoptotic threshold $w_a$ to be slightly lower at 3 \%.

The chemotactic coefficient of the endothelial cells was set to $\chi_m =2600$ cm$^2$ s$^{-1}$ M$^{-1}$ \cite{Anderson98}. The growth rate of the endothelial cells was estimated non-dimensionally by comparing it to the growth rate of the cancer cells. Endothelial cells are known to have much longer doubling times than cancer cells \cite{paweletz89}, and we therefore set $\alpha_m= \alpha_n/10\approx 0.07$. The matrix degradation rate $\beta_f$ has not yet been measured experimentally and we therefore resort to an estimate of this parameter. If we assume that a cancer cell degrades a volume of matrix which is of the same order of magnitude as the size of the cell itself during one cell cycle, then the degradation rate can be estimated to $\beta_f=10^{-13}$ cm$^3$ cell$^{-1}$ s$^{-1}$.

The diffusion coefficient of VEGF was set to $D_g=2.9 \times 10^{-7}$ cm$^2$ s$^{-1}$ \cite{Anderson98}. The production rate of VEGF by hypoxic cells was set to $\alpha_g=1.7\times10^{-22}$ mol cell$^{-1}$ s$^{-1}$, similar to the value used in \cite{jain07} (0.08 pg cell$^{-1}$ day$^{-1}$). The uptake of VEGF by endothelial cells has not been experimentally determined and we therefore estimate it to be of the same order as the production rate and set $\beta_g=\alpha_g=10$. All parameters in the model are summarized in Table \ref{core_param} in their non-dimensional and dimensional form, where appropriate.

\section{Results}\label{Res}
We have implemented our model \eqref{nutrient2}-\eqref{vegf2} using  \textsc{matlab} (version 7.1, The MathWorks, Inc., Natick, MA). We used the function {\tt pdepe}, which discretizes the equations in space to obtain a system of ordinary differential equations\footnote{The \textsc{matlab} codes will be available from the corresponding author upon request.}. The system of equations  \eqref{nutrient2}-\eqref{vegf2} is solved in the domain $\Omega=[0,1]\times(0,\tau_{max}]$, where $\tau_{max}$ will vary depending on the scenario we want to model. 

The model is supplied with the following initial conditions:
\begin{equation}\label{base_ini}
\begin{aligned}
w(\xi,0) &= 1.0, \qquad \xi \in [0,1], \\
n(\xi,0) &= 0.93\exp(-200\xi^2), \\
m(\xi,0) &= 0.01,\\
f(\xi,0) &=1-n(\xi,0)-m(\xi,0)-0.05,
\end{aligned}
\end{equation}
all other initial data being zero. We consider an initial avascular tumor located at $\xi=0$ in a tissue with a uniform background distribution of endothelial cells (blood vessels) and extracellular matrix. The oxygen concentration is initialized at its saturation value in the entire domain. We assume that all cells and chemical species remain within the domain, and therefore no-flux boundary conditions are imposed on $\partial \Omega$ for all equations. This is clearly a simplification of the actual situation, but it will already give instructive results (see \cite{ChaplainMatzavinos} for a similar situation). First we will present simulation results from the baseline version of the model and in the following sections we will modify the model to incorporate several modes of treatment.

\subsection{Baseline}
\begin{figure}[hb!]
\begin{center}
\includegraphics[width=10cm]{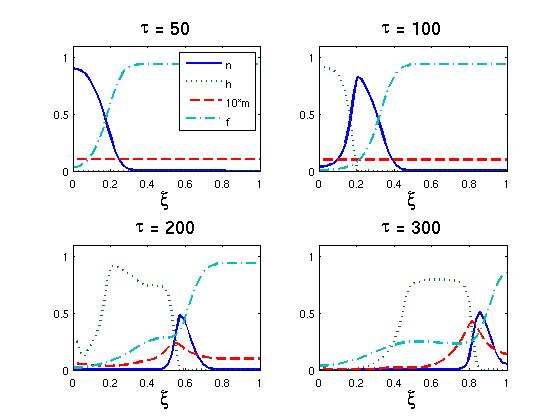}
\end{center}\caption{\label{fig:baseline_a}The spatial distribution of normoxic, hypoxic, endothelial cells and extracellular matrix at $\tau=50,100,200,300$ for the baseline scenario. Notice that the density of endothelial cells is multiplied by 10 for better visibility. }
\end{figure}

This case represents the baseline scenario when tumor growth only is limited by intrinsic constraints such as oxygen supply and the surrounding stromal tissue. The simulation lasted $\tau_{max}=300$ time steps (200 days) and the results of the simulation can be seen in Figure \ref{fig:baseline_a}, which shows the spatial distribution of the normoxic cells ($n$), hypoxic cells ($h$), endothelial cells ($m$) and the extracellular matrix ($f$) for four different time points of the simulation. Note that the density of the endothelial cells has been scaled up by a factor of 10 in all plots for better visualization.

At the beginning of the simulation we see the initial tumor growing and invading the surrounding tissue by degrading the extracellular matrix, but when the tumor reaches a critical size (approx.~$3\, mm$ or after $250$ division cycles) the oxygen supplied by the endothelial cells becomes insufficient and we observe the emergence of hypoxic cells in the core of the tumor (Figure \ref{fig:baseline_a}). The hypoxic cells secrete VEGF, which triggers endothelial proliferation and chemotaxis towards the hypoxic region. This consequently increases the oxygen production and leads to sustained tumor invasion. The end result is a traveling wave like scenario with a leading edge of normoxic cells followed by hypoxic cells and finally by a bulk of apoptotic cells (not shown). The density of endothelial cells has a peak at the edge of the tumor, is zero inside the tumor and remains approximately at the initial value in the surrounding tissue. The density of endothelial cells at the tumor-host interface is five-fold higher than the initial value, which implies that an angiogenic response has occurred due to the hypoxia experienced by the cancer cells.

The above scenario is firmly established as the typical scenario for early tumor growth \cite{sutherland88, plank03} and has also been reproduced in previous mathematical models \cite{orme96, alarcon06}, but as many of the parameters by necessity were estimated rather than experimentally determined we also investigated variations of these parameters within a reasonable range (data now shown). For example, increasing the oxygen consumption rate of the cells $\beta_w$ or the decay rate of oxygen $\gamma_w$ leads to an earlier appearance of hypoxia. Increasing the reproduction rate of endothelial cells gives rise to a stronger angiogenic response. A decrease of the diffusion constant of the angiogenic factor $D_g$ results in a more persistent gradient of angiogenic factor and a wider spread of endothelial cells, without, however, a greater number of normoxic tumor cells. This outcome is also seen if the chemotactic sensitivity $\chi_m$ of endothelial cells is increased. Future work will include a more systematic exploration of the parameter space and will indicate to experimentalists which parameters influence the tumor behavior most.

\subsection{Anti-angiogenic therapy}
Anti-angiogenic therapies are considered to be \textit{cytostatic} in the sense that the drugs used are not toxic to the cells, but instead inhibit some mechanism essential for cell division or a specific function. Probably the most well-known anti-angiogenic drug is Avastin (Bevacizumab) \cite{adams05}, which acts on endothelial cells by inhibiting the function of VEGF. The drug thus influences the endothelial cells' ability to react to the VEGF produced by hypoxic cells in the tumor and consequently is expected to reduce the angiogenic response and therefore inhibit further tumor invasion. 

We will model the effect of this type of drug by altering the behavior of the endothelial cells with respect to VEGF. The growth factor increases the proliferation and induces chemotactic movement of the endothelial cells, and we will model the effect of the drug by altering the parameters that correspond to these two processes. Avastin is usually administered intra-venously and therefore arrives at the tumor site through the blood vessels. As the endothelial cells in our model represent blood vessels these cells will be instantly affected by the drug, and there is therefore no need to introduce a new chemical species for the drug. Instead we can alter the values of the parameters under consideration directly depending on the concentration of the drug in the blood. Avastin is usually administered in 1-2 week intervals, but because the half-life of the drug is very long (1-2 weeks) \cite{bergsland04} the concentration of the drug in the blood can, as a first approximation, be considered constant. This implies that we can alter the parameter values at the start of the treatment and keep them constant during the entire treatment period. As mentioned before Avastin decreases the endothelial cell proliferation $\alpha_m$ and the chemotaxis coefficient $\chi_m$, we therefore model the effect of the drug by reducing these two parameter by a factor 10 during the treatment, i.e. when $\tau_s \leq \tau < \tau_{e}$ we let
\begin{equation*}
\begin{aligned}
\alpha_m & \rightarrow \alpha_m/10, \\
\chi_m & \rightarrow \chi_m/10, \\
\end{aligned}
\end{equation*}
where $\tau_s$ and $\tau_e$ are the start and end times of the treatment. In order to further investigate the effect of the drug we will decouple the inhibitory effect on VEGF and separately consider the impact of inhibiting the proliferation and the chemotaxis.

The approach we have used is of course a highly simplified way to model the treatment. A more complete model would include the pharmacokinetics and pharmacodynamics of the drug and also treat the effect of the drug on endothelial cells in more detail,  but our approach serves as a first attempt at incorporating the effects of an anti-angiogenic drug into a detailed spatial model of tumor-host interaction. 

\begin{figure}[c]
\centerline{\includegraphics[width=6cm]{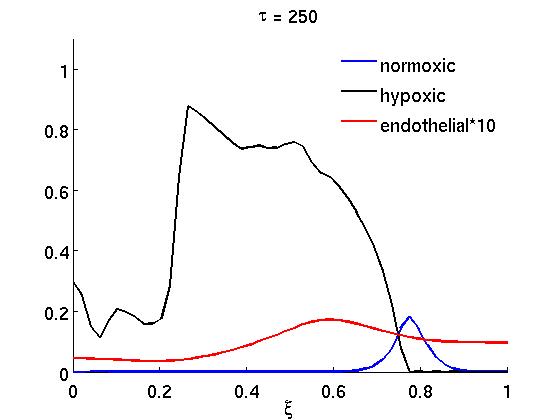}\includegraphics[width=6cm]{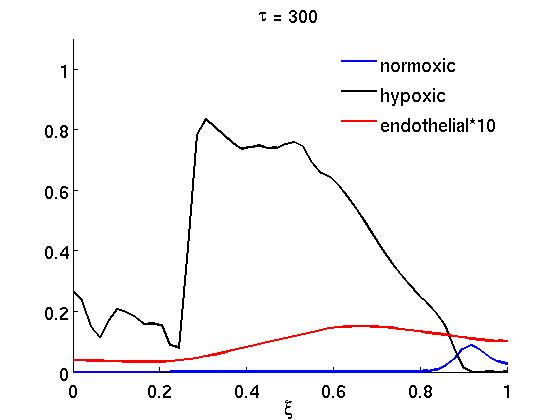}}
\caption{\label{fig:reduce_both} The spatial distribution of normoxic, hypoxic and endothelial cells during and after anti-angiogenic therapy. The proliferation $\alpha_m$ and the chemotaxis coefficient $\chi_m$ of the endothelial cells are reduced by a factor of $10$ while $200\le\tau\le 300$ (all other parameters are as in the baseline scenario). }
\end{figure}

The results of the anti-angiogenic drug on the system can been seen in Figure \ref{fig:reduce_both}, which shows the spatial distribution of normoxic, hypoxic and endothelial cells at (a) $\tau=250$ and (b) $\tau=300$. The drug was administered in the time interval $200 \leq \tau < 300$ (approx. 10 weeks) and in this simulation both effects of the drug were considered (i.e. both proliferation and chemotaxis). Up to the point where the treatment begins the dynamics are identical to that of the baseline scenario and are therefore not shown. When the treatment begins the endothelial cells become unresponsive to the VEGF and this has an obvious effect on the growth dynamics. The endothelial cells are now much more dispersed and not centered anymore at the leading edge of the hypoxic cell population (Figure \ref{fig:reduce_both}a). This effect is even more pronounced at the end of the treatment (Figure \ref{fig:reduce_both}b), when the endothelial cells are almost evenly distributed in the whole domain. This consequently decreases the oxygen supply to the normoxic tumor cells and leads to a lower cell density in the proliferating rim. The treatment could therefore be considered successful, but it should be noted that when the treatment ends the endothelial cells will regain their proliferative and chemotactic abilities and this could potentially lead to re-growth of the tumor.

\begin{figure}[c]
\centerline{\includegraphics[width=6cm]{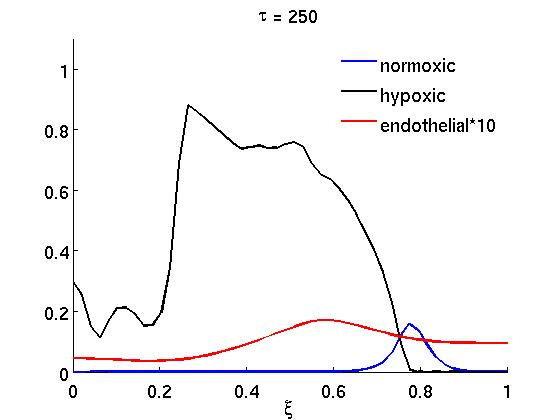}\includegraphics[width=6cm]{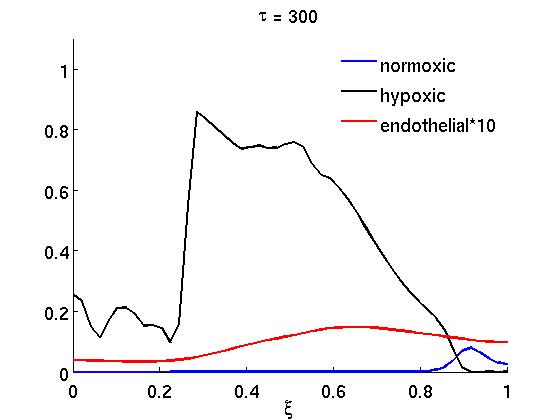}}
\caption{\label{fig:reduce_alpha}The spatial distribution of normoxic, hypoxic and endothelial cells during and after anti-angiogenic therapy when the proliferation $\alpha_m$ of the endothelial cells is reduced by a factor of $10$ while $200\le\tau\le 300$ (all other parameters are as in the baseline scenario).}
\end{figure}

\begin{figure}[c]
   \centerline{\includegraphics[width=6cm]{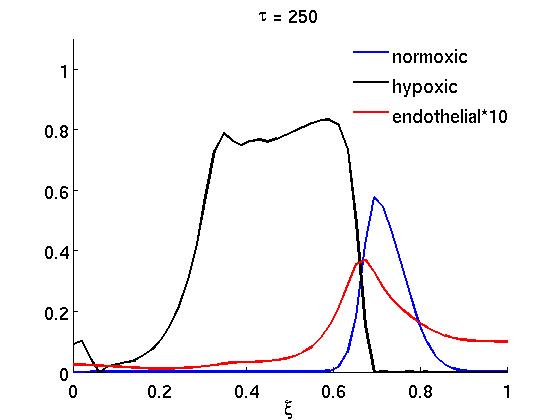}\includegraphics[width=6cm]{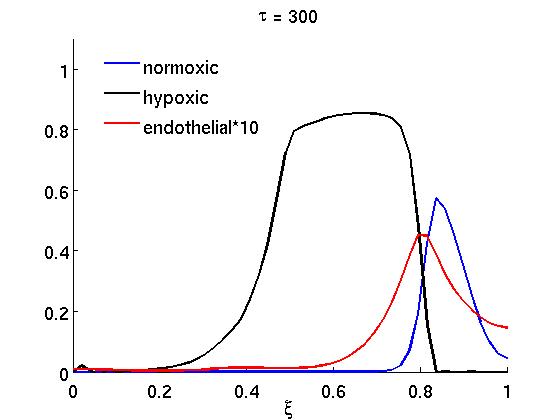}}
\caption{\label{fig:reduce_chi}The spatial distribution of normoxic, hypoxic and endothelial cells during and after anti-angiogenic therapy (at times $\tau=250$ and $\tau=300$) when the chemotaxis coefficient $\chi_m$ of the endothelial cells is reduced by a factor of $10$ while $200\le\tau\le 300$ (all other parameters are as in the baseline scenario).}
\end{figure}

Figure \ref{fig:reduce_chi} shows the result of the treatment if the anti-angiogenic drug only affects the endothelial cell chemotactic sensitivity $\chi_m$ (the duration of the treatment is as above). In this case the growth dynamics are quite similar to the baseline scenario, but on the other hand if only proliferation is affected (Figure \ref{fig:reduce_alpha}) the dynamics are similar to the full treatment. This suggests that inhibiting the proliferation of endothelial cells is the more important of the two effects the drug has.

\subsection{Modelling cytotoxic therapy}\label{cytotox_therapy}
We now expand our model to include therapy with a cytotoxic drug. The effects of this type of treatment has been investigated using several mathematical models \cite{panetta97,afenya96,abundo89,martin90,usher96,Ledzewicz05,Swierniak}. These studies have mostly focused on the efficacy of specific drugs, but we will instead model the effects of a general cytotoxic drug that affects cells in the proliferative state (i.e. normoxic cells), and investigate how the drug characteristics influence the outcome of the drug therapy. Many drugs used in cytotoxic chemotherapy, for example spindle poisons (paclitaxel) and drugs that interfere with the DNA synthesis (doxorubicin, fluorouracil) affect only proliferating cells. The concentration of the drug $c$ is introduced as a new variable and it is described by a reaction-diffusion equation similar to that of oxygen,
\begin{equation}\label{drug}
\frac{\partial c}{\partial \tau}(\xi,\tau) =  D_c \frac{\partial^2 c}{\partial \xi^2} +\alpha_c(\tau)m(1-c) - \gamma_cc -  k \gamma_n nc.
\end{equation}
Notice that this equation is already in non-dimensional form and that $0\le c\le 1$. The drug is assumed to be delivered through blood infusion and therefore enters the tissue from the blood stream. The production rate of the drug is therefore proportional to the density of endothelial cells, which implies that regions with higher vascular density will experience a higher concentration of the drug. Instead of looking at a constant drug supply we include a scheduling of the drug by letting the production rate $\alpha_c(\tau)$ be a time-dependent function, where $\alpha_c(\tau)>0$ corresponds to the drug being delivered to the tissue. This is of course a highly simplified picture of the true dynamics of drug delivery. A more realistic approach would be to also include the pharmacokinetics of the drug into the model \cite{bonate2006}, but as we will model a general cytotoxic drug we will use this simplified form and only consider the dynamics of the drug at the tumor site. The drug diffuses (for simplicity we assume the same diffusion constant $D_c$ as for oxygen) in the tissue and decays at a constant rate $\gamma_c$. It affects the normoxic cells alone and is consumed at a rate $k \gamma_n$ when it kills normoxic cells. The normoxic cells are driven into apoptosis at a rate proportional to the drug concentration
\begin{equation}
\frac{\partial n}{\partial \tau}(\xi,\tau) = \dots -\gamma_nnc,
\end{equation} 
with rate constant $\gamma_n>0$. The dimensionless parameter k corresponds to the amount of drug needed to kill one unit volume of proliferating cells. For sake of simplicity, we have worked with $k=1$. Finally, to balance the loss of normoxic cells, the equation for apoptotic cells is amended according to
\begin{equation}
\frac{\partial a}{\partial \tau}(\xi,\tau) = \dots +\gamma_nnc.
\end{equation}

\begin{figure}[c]
   \centerline{\includegraphics[width=6cm]{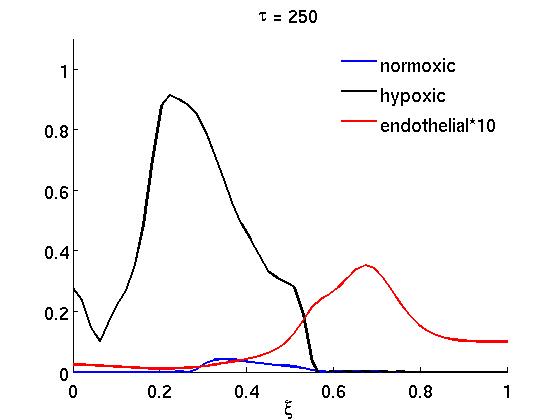}\includegraphics[width=6cm]{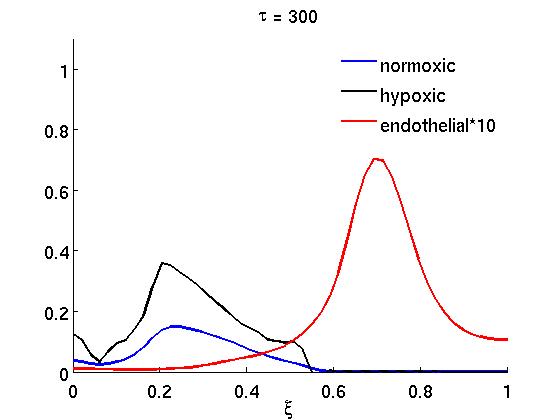}}
\caption{\label{fig:chemo1}The spatial distribution of normoxic, hypoxic and endothelial cells after chemotherapy has been applied with the drug parameters $\gamma_n=10$ and $\gamma_c=0.1$ at times  $\tau=250$ and $\tau=300$.}
\end{figure}

\begin{figure}[c]
   \centerline{\includegraphics[width=6cm]{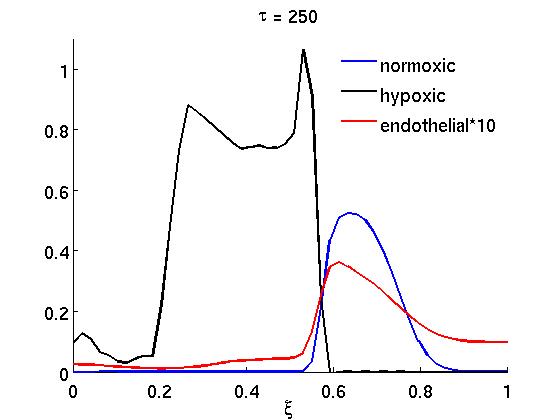}\includegraphics[width=6cm]{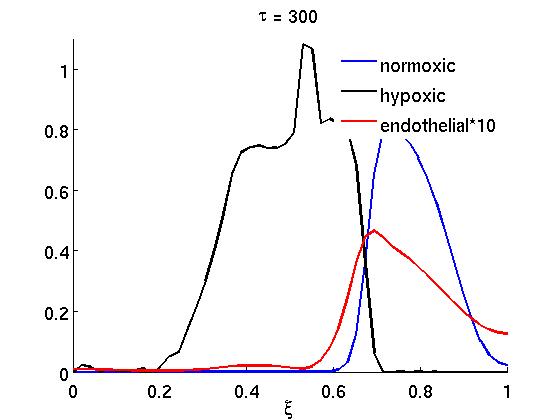}}
\caption{\label{fig:chemo2}The spatial distribution of normoxic, hypoxic and endothelial cells after chemotherapy has been applied with the drug parameters $\gamma_n=1$ and $\gamma_c=1$ at times  $\tau=250$ and $\tau=300$.}
\end{figure}

\begin{figure}[c]
   \centerline{\includegraphics[width=13.5cm]{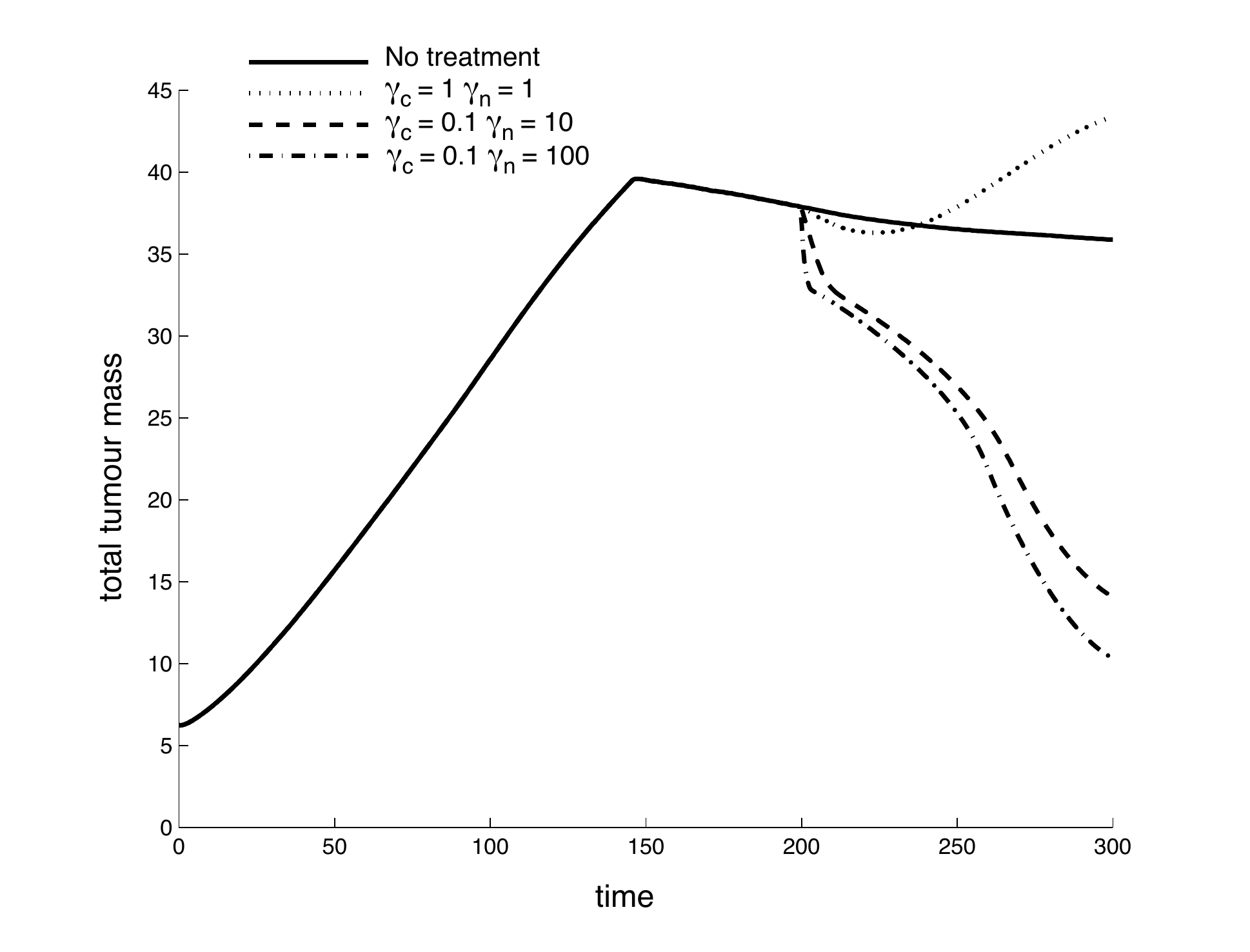}}
\caption{\label{fig:mass}The time evolution of the total tumor mass (normoxic plus hypoxic cells) for the baseline scenario and three different chemotherapy treatments. It can be observed that chemotherapy not only has the capability to decrease, but also to increase the growth rate of the tumor. The decrease in tumor mass prior to treatment is due to the emergence of apoptotic cells in the center of the tumor and the linear growth rate would be preserved if the apoptotic cells were included in the total tumor mass.}
\end{figure}

The treatment schedule was fixed at
\begin{equation*}
\alpha_c(\tau)=100\sum_{k=0}^5 \exp(-4(\tau-(200+2k))^2),
\end{equation*}
which corresponds to a treatment that starts at $\tau=200$ and consists of 6 infusions of the drug. Each infusion lasts approximately 1 cell cycle, and the interval between two infusion is 2 cell cycles. This treatment schedule might not be realistic for some drugs (e.g. due to toxicity), but in order to separate the effects of scheduling and drug characteristics we will use this basic schedule in our investigation.
An application of this treatment schedule with parameters $\gamma_n=10$ and $\gamma_c=0.1$ can be seen in Figure \ref{fig:chemo1}, which shows the post-treatment density of normoxic, hypoxic and endothelial cells at $\tau = 250$. In this case the treatment is successful and the tumor mass is significantly reduced compared to the untreated baseline scenario (Figure \ref{fig:baseline_a}). From this plot we can also observe that the reduction in normoxic cells is largest where the endothelial cells are located and that the surviving fraction of cancer cells now reside in a region of low vascular density. For other values of the drug parameters, in particular a reduced kill rate $\gamma_n$, the treatment is less successful, and it can even lead to enhanced tumor growth, as can be seen in Figure \ref{fig:chemo2}. This plot shows the post-treatment density of cells (at $\tau = 250$), and comparing it to the baseline simulation we can observe that the total number of normoxic cells is actually higher although chemotherapy has been applied. This is further illustrated in Figure \ref{fig:mass}, which shows the time evolution of the total tumor mass (normoxic plus hypoxic) for the baseline scenario and three different drug characteristics. For $\tau < 200$ the dynamics are identical as the initial conditions and governing equations are identical, but when the therapy is applied at $\tau=200$ they diverge. In the initial growth stage ($\tau < 150$) we observe an approximately linear increase in tumor mass. This is followed by a sharp transition to a slowly decreasing tumor mass which is due to the build up of apoptotic cells in the center of the tumor (if the apoptotic cells are included in the tumor mass the linear growth rate is preserved). When the treatment is applied the dynamics clearly depend on the drug parameters and we observe that the treatment can decrease, but also increase the total tumor mass.

In order to investigate this further and to fully characterize the impact of the kill and decay rate on the efficacy of the treatment we performed systematic measurements of the post-treatment tumor mass in the range [1,100] $\times$ [1,100] of $\gamma_c$ and $\gamma_n$. This was done by running the simulation for 400 evenly distributed points in the parameter range and at $\tau_s=300$ measure the total mass of the tumor (normoxic plus hypoxic cells). The tumor mass was then normalized with respect to the untreated case. The normalized post-treatment mass is thus given by,
\begin{equation}\label{normalized_mass}
M=\frac{1}{M_0}\int_0^1 \left(n(\xi,\tau_s)+h(\xi,\tau_s)\right) \ud\xi,
\end{equation}
where $M_0$ is the untreated tumor mass. 

\begin{figure}[c]
   \centerline{\includegraphics[width=15cm]{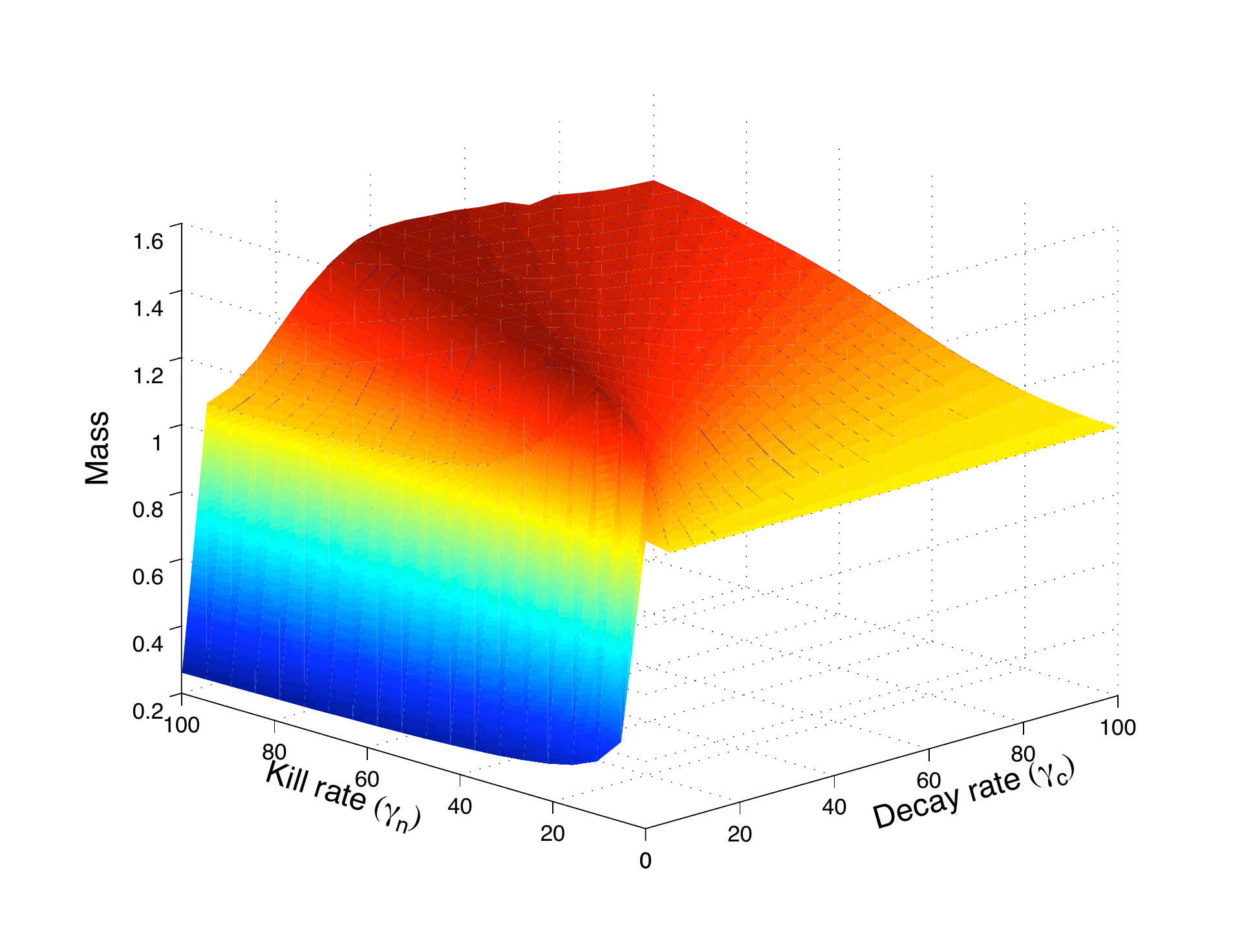}}
\caption{\label{fig:chemomass}Post-treatment normalized tumor mass as a function of the kill rate ($\gamma_n$) and decay rate ($\gamma_c$). The untreated case corresponds to $M=1$.}
\end{figure}

\begin{figure}[c]
   \centerline{\includegraphics[width=15cm]{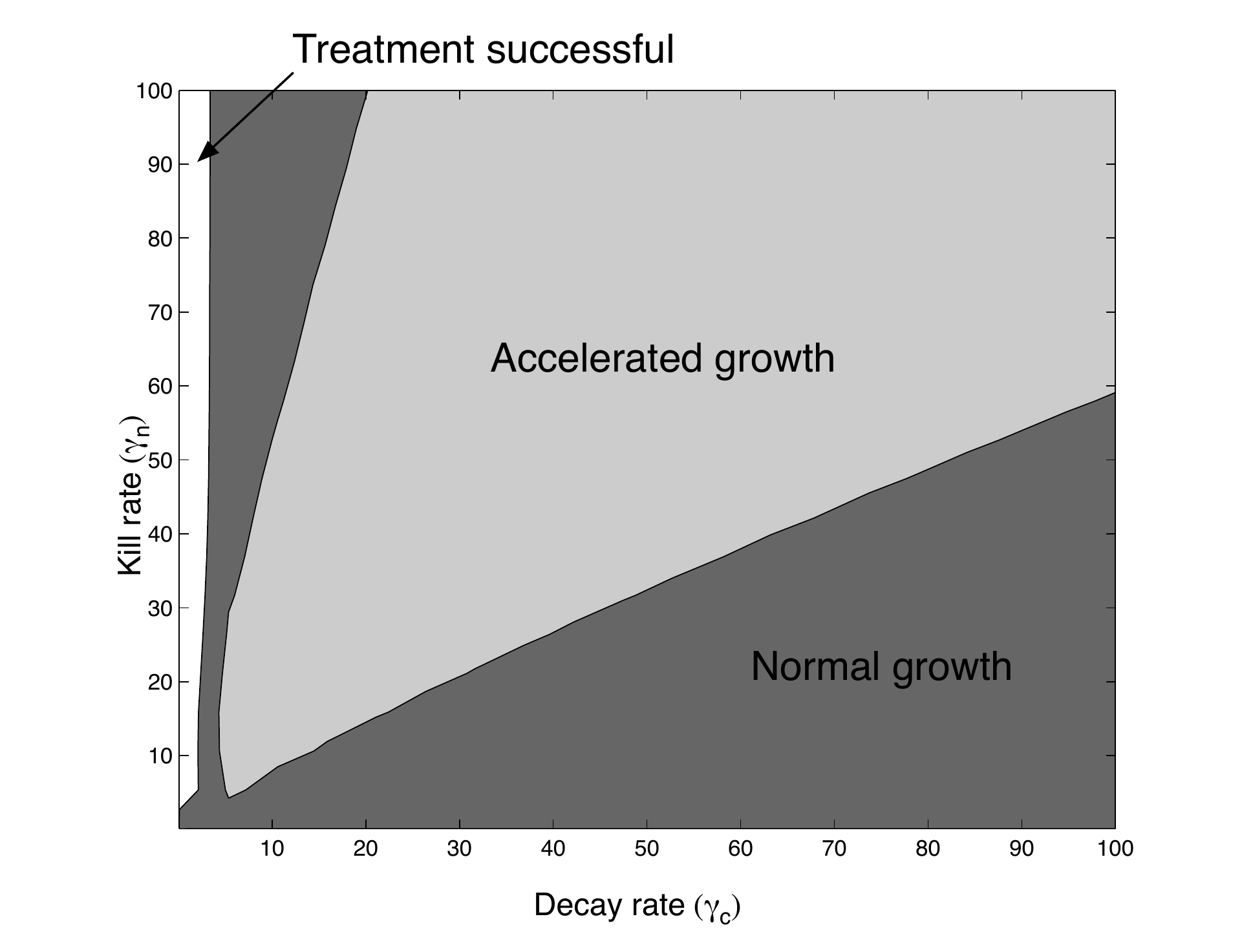}}
\caption{\label{fig:chemotreat}The division of drug parameter space into three distinct regions: (i) successful treatment (ii) normal growth and (iii) accelerated growth. Here successful treatment is defined by a normalized tumor mass $M\le 0.75$ and accelerated growth by a normalized tumor mass $M\ge 1.2$ (see equation \eqref{normalized_mass} for the definition of $M$).}
\end{figure}

The results can be seen in Figure \ref{fig:chemomass} and show that in a significant part of the parameter space the treatment actually increases the tumor mass. We can observe that the treatment is most successful for high kill rates and low decay rates, where the post-treatment tumor mass $M \approx 0.2$. On the other hand for intermediate kill and decay rates the tumor mass is higher than in the untreated case reaching a post-treatment mass of $M \approx 1.5$. For low kill rate and high decay rates we have $M \approx 1$, which is expected as the untreated case corresponds to $\gamma_n$=0. The parameter space can therefore be divided into three distinct regions: (i) Successful treatment ($M\le 0.75$), (ii) Unaffected growth, and (iii) Accelerated growth ($M\ge1.2$). This is illustrated in Figure \ref{fig:chemotreat}, which shows that a large region in the parameter space in fact gives rise to accelerated growth.

\section{Discussion}\label{Discussion}

The model presented here is similar to previous reaction-diffusion models of tumor invasion, but also contains a number of novel features that are worth mentioning. Firstly we have included density driven motion of the cancer cells, which acts by pushing cells out from regions of high density. This mechanism is potentially important as tumor tissue is known to be highly irregular and to have higher cell density than normal tissue. Although we have not seen any obvious effect of this mechanism, this might be different when considering the system in two dimensions, where this type of density driven motion is known to create front instabilities \cite{sarloos02}, which can cause fingering protrusions. Our approach of using a degenerate diffusion term to account for cell crowding is simpler and more transparent than a volume fraction formulation with a force balancing term. This approach also meshes more easily with standard representations of haptotaxis and random motility.

Secondly we have taken the extracellular matrix into account when considering the maximum density of cells. This implies that the cancer cells cannot move into regions where the matrix density is too high and thus imposes a growth constraint which clearly is present in real tissue. Further it implies that degrading the matrix does not only induce haptotaxis, through the establishment of matrix gradients, but also pure random movement, as the degradation frees up space that the cancer cells can move into.

In the basic setting of our model, which represents unconstrained tumor growth and invasion, the simulations produce results that are in qualitative agreement with the typical scenario of early tumor growth: the tumor starts growing as an avascular tumor that reaches a critical diffusion limited size of a couple of millimeters. At this stage the tumor has outgrown its nutrient supply and cells in the center of the tumor become hypoxic due to the low oxygen concentrations. This triggers them to secrete VEGF which diffuses in the tissue and stimulates endothelial cells to form blood vessels towards the hypoxic site through proliferation and chemotaxis. As a result of this angiogenic response the tumor can continue to grow and invade the surrounding tissue. 

Although this does not give any new insight into the dynamics of tumor invasion it nevertheless gives us a reference with which we can compare the different treatment strategies we apply. This is important as it gives us the possibility to examine different types of treatment within a fully developed framework of tumor growth. 

The results from the simulations of anti-angiogenic treatment show that it has the capability to reduce tumor mass and slow down tumor invasion. This is in qualitative agreement with experimental results \cite{gerber05}, but the strength of modelling this treatment is that we have the capability to decouple the effects of the drug on endothelial cells. In doing so we observe that proliferation of endothelial cells seems to be more important in establishing sufficient angiogenesis than chemotaxis of endothelial cells. Reducing the chemotaxis alone seems to have little or no therapeutic effect, while reducing the proliferation of the endothelial cells has a similar effect as the full treatment. This is an interesting observation and suggests that drugs that specifically target  proliferation of the endothelial cells could be more efficient as anti-angiogenic agents.

In the case of the cytotoxic treatment we also made observations that are surprising and even counterintuitive. The fact that the cytotoxic treatment leads to accelerated tumor growth is quite unexpected, as even a drug with a low kill rate will reduce the number of normoxic cells in the tumor, and from a naive view-point this would reduce the tumor mass. The fact that we observe the opposite, an increase in the tumor mass, reveals that the dynamics of the growth are more complex than one would expect. The results suggests that there exists a complicated interplay between the different cell types occupying the tumor and that disrupting the balance between these can have unexpected consequences. 

When the kill rate of the drug is low (or the decay rate is high) only a small fraction of the normoxic cells at the boundary of the tumor are driven into apoptosis. As the reduction in number of normoxic cells, at the tumor boundary, lowers the total oxygen consumption this leads to a conversion of previously hypoxic into normoxic cells. The interesting fact is that the number of normoxic cells after treatment is actually higher than before, which suggests that the local carrying capacity is increased in the accelerated growth regime. This can only occur if the local density of endothelial cells is increased, which implies that the treatment indirectly influences the vascular density at the tumor-host interface. This also highlights that cytotoxic therapy can have indirect effects on the micro-environment of the tumor and can even make the growth conditions more beneficial for the tumor. This observation might also shed new light on the efficacy of existing cytotoxic drugs and possibly guide the development of new drugs. It should be noted that a model used for this specific purpose would need to be properly experimentally parametrized and validated, however, at least the model presented in this paper is a first attempt at modelling the influence of tumor-host interactions on chemotherapy. The fact that the models takes into account the interactions between the tumour and the stromal cells means that the results need to be compared with in vivo rather than in vitro studies. One possibility is to compare tumor growth rates during cancer therapy with different drugs under the same delivery schedules. The decay/kill rates of these drugs could also be measured, which would make a comparison with the model results possible.

The model originally contained further variables which we later disregarded for the sake of simplicity. In future extensions of this work we will consider different types of extracellular matrix (namely tissue matrix and clotting matrix) as well as reactive cells. Reactive cells are attracted by gradients of clotting matrix and can turn clotting matrix into tissue matrix. Eventually one may observe an encapsulation of the tumor by fibrous tissue. Another simplification that was made in the present model is that matrix degrading enzymes are located in the cell membranes of normoxic cells alone. In future work we plan to include diffusible matrix degrading enzymes as a new dependent variable. Finally, we plan to take the numerical simulations to two-dimensional domains to allow for greater morphological complexity.

\section{Conclusion}\label{Conclusion}
\begin{sloppypar}
We have seen in the present paper that mathematical models with a spatial component allow for complex and sometimes counterintuitive growth dynamics of tumors. Therefore, mathematical models serve as an ideal tool to evaluate the efficacy of cancer drugs. An interface between experimental biologists and mathematicians can streamline this process considerably. 
\end{sloppypar}

We built our model based on few basic assumptions summarized as follows: (i) initial tumor growth is only constrained by the surrounding tissue, (ii) the tumor mass rapidly outgrows nutrient supply, (iii) subsequent tumor cell death stimulates stromal response, (iv) new stroma develops vasculature and (v) tumor growth resumes. These assumptions are well supported by available literature \cite{sutherland88, plank03}. However, they are only a partial set of variables that one may consider in building a mathematical model of tumor growth. They were chosen because in our view they are a minimal set that can realistically capture the effects of cytotoxic and cytostatic drugs. Naturally, empirically determined parameters would have to be introduced in order to make the model entirely realistic. Nonetheless, there are novel features as well as conclusions of the model that are worth reporting at this stage.

One novel feature of this model is the way tumor cell migration is modeled, i.e., tumor cells are driven by crowding, in addition to the more conventional unbiased diffusion and haptotaxis. This introduces a measure of realism because it is generally recognized that tumor invasion is partly due to a pressure build-up that pushes cells into the surrounding tissue. Further we have also taken into account the extracellular matrix when considering the maximum cell density in the tissue, which implies that the stroma restrains the growth of the tumor.  

Perhaps the most compelling novel feature of our model is chemotherapeutic drug delivery by a spatial source, i.e., endothelial cells.
Cancer chemotherapy has been the subject of countless modelling efforts \cite{abundo89, costa95, usher96, gaffney04, hinow07, barbolosi2001, murray90, Ribba, panetta2003}. However, spatial aspects of chemotherapeutic drug delivery seem to have received less attention so far. This is probably because the main objective is to evaluate the ability of a drug to debulk a tumor by inhibiting growth kinetics or killing cells. In contrast, in our model we consider the effects of removing cells (by drug killing or growth inhibition) in specific locations of a growing tumor. That is, we view chemotherapeutic drugs as agents that have a localized spatial effect on the growth of the tumor. This means that the endothelial cells have a dual role in our model. They supply the tumor with essential nutrients, but also when therapy is applied, with detrimental drugs. The angiogenic response therefore becomes crucial as it is necessary for continued growth, but also is utilized for drug delivery, and again this highlights the importance of taking into account the complex dynamics of tumor-host interactions when modelling drug therapy.

\section*{Acknowledgments}
We would like to thank Walter Georgescu, Patrick Reed, Glenn F. Webb and Candice Weiner for their participation in the creation of the model. We thank the VICBC for hosting the annual workshop, in particular, we express our thanks to the scientific organizer Lourdes Estrada and to Yolanda Miller. Funding from the National Institutes of Health (Grant number: U54 CA 11307) is greatly appreciated. PH is supported by an IMA postdoctoral fellowship and BPA and JMG were partially supported by the NSF under award DMS-0609854. We thank two anonymous referees for their careful reading of the manuscript and helpful remarks.


\section*{Appendix}
The length is re-scaled with the typical size of an early stage tumor $L=1$ cm and time with $T=16$ h the typical time of a cell cycle for cancer cells \cite{calabresi93}. The cell densities are re-scaled with a maximum cell density of $v_{max}=10^{-8}$ cell/cm$^3$, which corresponds to each cell having a diameter of approximately $20\ \mu$m \cite{Casciari92}. The oxygen concentration is re-scaled by the background concentration $w_{max}=6.7 \times 10^{-6}$ mol/cm$^3$ \cite{Anderson05} and the VEGF concentration by the maximal concentration $g_0=10^{-13}$ mol/cm$^3$ \cite{Anderson98}. 

We scale the independent variables
\begin{equation*}
\xi = \frac{x}{L}, \quad \tau = \frac{t}{T}.
\end{equation*}
The dependent variables are scaled as follows
\begin{align*}
\tilde{w}&= \frac{w}{w_{max}}, \quad \tilde{n} = \frac{n}{v_{max}}, \quad
\tilde{h}= \frac{h}{v_{max}}, \quad \tilde{a} = \frac{a}{v_{max}}, \\
\tilde{m}&= \frac{m}{v_{max}}, \quad
\tilde{f}= \frac{f}{v_{max}}, \quad \quad \tilde{g} = \frac{g}{g_{tot}}.
\end{align*}
Notice that the concentration of extracellular matrix $f$ has the unit cells$/cm^3$. To convert this into a concentration of the unit moles $/cm^3$ one has to multiply with a conversion factor of the unit moles/cell.  This conversion factor can depend on the type of matrix molecules and the type of cells considered. Equation \eqref{nutrient} for the oxygen becomes
\begin{equation*}
\frac{\partial \tilde{w}}{\partial \tau}(\xi,\tau) =  \frac{TD_w}{L^2} \frac{\partial^2 \tilde{w}}{\partial \xi^2} +\alpha_wTv_{max}\tilde{m} (1-\tilde{w}) - \frac{\beta_wTv_{max}}{w_{max}}(\tilde{n}+\tilde{h}+\tilde{m})\tilde{w} - \gamma_w T\tilde{w}.
\end{equation*}
We define new dimensionless parameters
\begin{equation*}
\tilde{D}_w =  \frac{TD_w}{L^2},\quad 
\tilde{\alpha}_w = \alpha_wTv_{max},\quad 
\tilde{\beta}_w = \frac{\beta_wTv_{max}}{w_{max}},\quad 
\tilde{\gamma}_w = \gamma_w T.
\end{equation*}
Upon dropping the tildes everywhere for notational convenience, our new dimensionless equation becomes
\begin{equation*}
\frac{\partial w}{\partial \tau}(\xi,\tau) =  D_w \frac{\partial^2 w}{\partial \xi^2} +\alpha_wm(1-w)- \beta_w(n+h+m)w - \gamma_ww.
\end{equation*}
A similar procedure was applied to the remaining equations \eqref{normoxic}-\eqref{vegf} and by dropping the tildes they return to their original form with the only difference that the parameters now are in their non-dimensional form. The complete simplified non-dimensional system that we will solve numerically in the following sections is as follows,
\begin{equation}\label{nutrient2}
\frac{\partial w}{\partial t}(x,t) =  D_w \frac{\partial^2 w}{\partial x^2} +\alpha_w m (1-w) - \beta_w(n+h+m)w - \gamma_w w,
\end{equation}

\begin{align}
\frac{\partial n}{\partial t}(x,t)&=  \frac{\partial }{\partial x}\left( (D_n\max\{n-v_c,0\}+D_m)\frac{\partial n}{\partial x}\right) - \frac{\partial }{\partial x}\left(\chi_n n\frac{\partial f}{\partial x} \right)  \label{normoxic2} \\
&\quad+\alpha_nn(v_{max}-v) -\alpha_h\HH(w_h-w)n + \frac{1}{10}\alpha_h\HH(w-w_h)h, \nonumber
\end{align}
\begin{equation}\label{hypoxic2}
\frac{\partial h}{\partial t}(x,t)= \alpha_h\mathcal{H}(w_h-w)n - \frac{1}{10}\alpha_h\mathcal{H}(w-w_h)h
-\beta_h\mathcal{H}(w_a-w)h,
\end{equation}
\begin{equation}\label{apoptotic2}
\frac{\partial a}{\partial t}(x,t)= \beta_h\mathcal{H}(w_a-w)h,
\end{equation}
\begin{equation} \label{endo2}
\frac{\partial m}{\partial t}(x,t) =  \frac{\partial }{\partial x}\left( D_m\frac{\partial m}{\partial x} -m \chi_m \frac{\partial g}{\partial x}\right) + \alpha_m mg(v_{max}-v), 
\end{equation}
\begin{equation}\label{tissue2}
\frac{\partial f}{\partial t}(x,t)=  - \beta_f n f,
\end{equation}
\begin{equation}\label{vegf2}
\frac{\partial g}{\partial t}(x,t)= D_g\frac{\partial^2 g }{\partial x^2}+ \alpha_g h - \beta_g m g.
\end{equation}
\section*{Figures and Tables}
\vspace{\baselineskip}


\clearpage               

\begin{table}
	\begin{center}
	\caption{\label{dependentVar} Variables of the mathematical model}
		\begin{tabular}{|l|l|}
			\hline
			Notation &  Concentration of \dots \\
			\hline
			$w$ & nutrient \\
			$n$ & normoxic cells \\
			$h$ & hypoxic cells \\
			$a$ & apoptotic cells \\
			$m$ & endothelial cells \\
			$f$ & extracellular matrix \\
			$g$ & VEGF\\
			$c$ & cytotoxic drug \\
			\hline 
		\end{tabular}
	\end{center}
\end{table}

\begin{table} 
	\begin{center}
	\caption{The scaling factors for the nondimensionalization procedure.\label{tab_scaling}}
		\begin{tabular}{|l|l|l|}
			\hline
			Parameter &  Value & Meaning\\
			\hline
			$T$ & 16 h  & cell cycle time \\
			$L$ & 1 cm & typical length scale \\
			$v_{max}$ & $10^8$ cells$/cm^3$ & maximum density of cells 	\\		
			$w_{max}$ & $6.7 \times 10^{-6}$ moles $/cm^3$ & saturation level of oxygen \\
			$g_{tot}$ & $10^{-13}$ moles $/cm^3$ & maximal concentration of VEGF\\
			\hline
 		\end{tabular}
	\end{center}
\end{table}

\begin{table}
	\begin{center}
		\caption{Model parameter values in both their nondimensional and dimensional forms.}\label{core_param}
		\begin{tabular}{|l|l|l|l|}\hline
			Parameter & ND-value & D-value & Reference\\
			\hline
			$D_w$ & $0.58$  & $10^{-5}$ cm$^2$ s$^{-1}$ & \cite{Anderson05}\\ 
			$\beta_w$ & $0.57$  & $6.25\times 10^{-17}\,$mol cell$^{-1}\,s^{-1}$ & \cite{Casciari92}\\ 	
			$\alpha_w$ & $1$ & - & -\\ 
			$\gamma_w$ & $0.025$ & - & \cite{Anderson05} \\ 
			$D_m$ & $5.8\times10^{-5}$ & $10^{-9}$ cm$^2$ s$^{-1}$ & \cite{Bray} \\ 
			$D_n$ & $5.8\times10^{-5}$ & - & \cite{Bray} \\ 
			$v_c$ & $0.8$ & - & -\\ 
			$\chi_n$ & $1.4 \times 10^{-4}$ & - & \cite{Anderson05} \\ 
			$\alpha_n$ & $\log 2$ & - & -\\ 
			$\alpha_h$ & $1.6$ & $2.8 \times 10^{-5}$ s$^{-1}$ & -\\
			$\beta_h$ & $0.32$  & $5.6\times10^{-6}$ s$^{-1}$ & \cite{borutaite05}\\
			$w_h$ & $0.05$  & $3.4\times10^{-7}$ mol cm$^{-3}$ & \cite{brown04}\\
			$w_a$ & $0.03$ & $2.0\times10^{-7}$ mol cm$^{-3}$ & \cite{brown04}\\
			$\chi_m$ & $2.1\times10^{-6}$ & $2.6 \times 10^3$ cm$^2$ s$^{-1}$ M$^{-1}$ & \cite{Stokes91}\\
			$\alpha_m$ & $0.07$ & - & \cite{paweletz89}\\ 
			$\beta_f$ & $0.5$ & $10^{-13}$ cm$^3$ cell$^{-1}$ s$^{-1}$ & -\\   			
			$D_g$ & $0.02$ & $2.9\times10^{-7}$ cm$^2$ s$^{-1}$ & \cite{Sherratt90}\\ 
			$\alpha_g$ & $10$ & $1.7\times10^{-22}$ mol cell$^{-1}$ s$^{-1}$ & \cite{jain07}\\
			$\beta_g$ & $10$ & - & -\\
			\hline 
 		\end{tabular}
	\end{center}
\end{table}
\clearpage

%

\end{document}